\documentclass{ifacconf}

\usepackage{graphicx}      
\usepackage{natbib}        
\usepackage{graphics}			
\usepackage{epsfig} 			
\usepackage{mathptmx} 			
\usepackage{times} 				
\usepackage{amsmath} 			
\usepackage{amssymb}  			
\usepackage{xcolor}
\usepackage{multirow}
\usepackage{amsmath}

\usepackage{tikz}
\newcommand\copyrighttext{%
  \centering\footnotesize Accepted for publication at the IFAC World Congress 2023, July 9-14, 2023, Yokohama, Japan. \\
\copyright 2023 the authors. This work has been accepted to IFAC for publication under a Creative Commons Licence CC-BY-NC-ND. }
\newcommand\copyrightnotice{%
\begin{tikzpicture}[remember picture,overlay]
\node[anchor=south,yshift=-10pt] at (current page.north) {\fbox{\parbox{\dimexpr\textwidth-\fboxsep-\fboxrule\relax}{\copyrighttext}}};
\end{tikzpicture}%
}

\begin{document}
\begin{frontmatter}

\copyrightnotice

\vspace{-10pt}

\title{Virtual-bike emulation in a series-parallel human-powered electric bike} 


\author{Stefano Radrizzani,} 
\author{Giulio Panzani,}
\author{Sergio M. Savaresi}

\address{All the authors are with the Dipartimento di Elettronica, Informazione e Bioingegneria, Politecnico di Milano, Via Ponzio 34/5, 20133 Milan, Italy. \\ (e-mail: stefano.radrizzani@polimi.it)}

\begin{abstract}               
Combining the advantages of standard bicycles and electrified vehicles, electric bikes (e-Bikes) are promising vehicles to reduce emission and traffic. The current literature on e-Bikes ranges from works on the energy management to the vehicle control to properly govern the human-vehicle interaction. This last point is fundamental in chain-less series bikes, where the link between the human and the vehicle behavior is only given by a control law. In this work, we address this problem in a series-parallel bike. In particular, we provide an extension of the virtual-chain concept, born for series bikes, and then we improve it developing a virtual-bike framework. Experimental results are used to validate the effectiveness of the solutions, when the cyclist is actually riding the bike.


\end{abstract}

\begin{keyword}
e-Bikes, virtual-chain, virtual-bike, vehicle control
\end{keyword}

\end{frontmatter}
\section{Introduction}
Electric bikes (e-Bikes) are fully-fledged human-powered hybrid vehicles, since the human and battery power are both involved in the vehicle propulsion. e-Bikes are a promising way to reduce emissions and traffic, combining the advantages of standard bicycles (small footprint and low emissions) and electric vehicles (reduced physical effort optimizing human efficiency) [\cite{alli2010epac,corno2017senza}].\\
Similar to other hybrid vehicles, there exist different e-Bikes architectures. Among them, the most common and studied e-Bikes are the parallel ones, i.e. traditional chained bikes, where the human power sums with the one coming from an electric motor before being mechanically transmitted to the wheel. On the other hand, in series bikes, which constitute another possible architecture, the human power is converted, through a generator mounted on the pedals, into electric power stored in a battery and then delivered to the wheel through an in-wheel electric motor. Therefore, the battery acts as an electrical transmission; hence, this architecture does not need a mechanical transmission and so series bikes are chain-less.\\
The bike market, and the current state of the art as well, is mainly focused on parallel bikes. In fact, series bikes require two electrical machines instead only one in parallel bike; moreover, they need a proper vehicle control law, due to the absence of the mechanical chain. For this last reason, the current literature on series bikes not only provides works related to energy management, e.g. \cite{guanetti2017optimal}, but also works aiming at properly linking the vehicle dynamics with the human behavior, as shown by \cite{corno2015senza,corno2017senza}. Particularly, \cite{corno2017senza} explored the potential of a chain-less bike by virtually emulating a chain transmission, whose chain ratio can be freely designed by the rider. This approach is named \textit{virtual-chain}.\\
Finally, there exists also series-parallel bikes, e.g. the prototype proposed by \cite{ongun2020hybrid2}, that  combine the different architectures.  

\begin{figure}[b]
\centering
\includegraphics[width = 0.66\columnwidth]{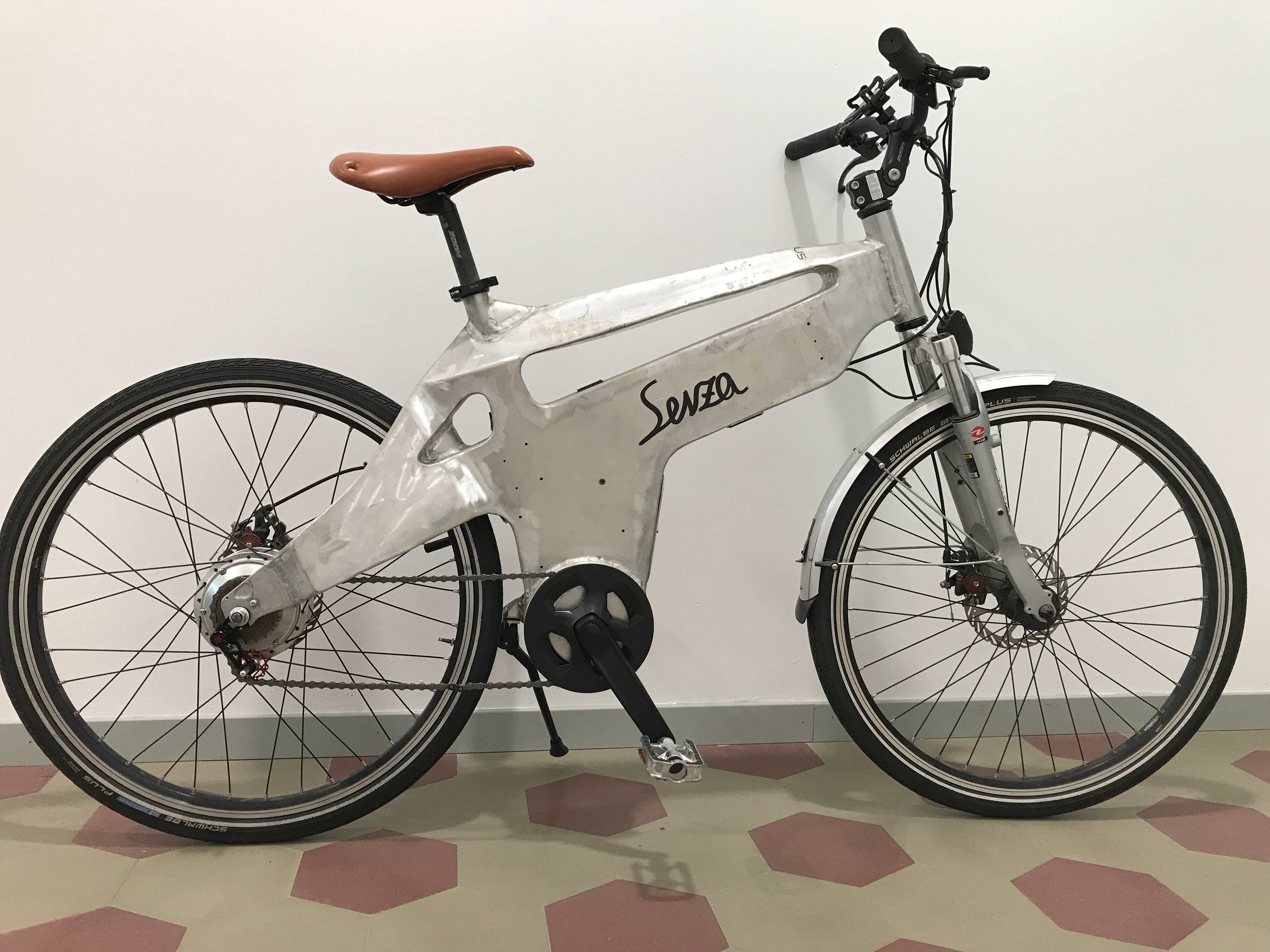}
\caption{Experimental seutp.}
\label{fig_senza}
\end{figure}

In this work, we consider the series-parallel bike in Fig. \ref{fig_senza}, which is the evolution of the series bike introduced in \cite{corno2015senza,corno2017senza}, in order to keep the advantages of the virtual-chain control strategy, overcoming the drawbacks of series bikes at low speed, generated by the absence of a chain. In the considered prototype, the parallel architecture is realized by adding mechanical chain, while the series one is activated by a free-wheel mechanism which is able to disengage the chain.

A preliminary contribution of this works is the extension of the virtual-chain control strategy for series bike to series-parallel ones. In particular, a switching control law is designed as a function of teh currectly active architecture, parallel or series.  However, the main contribution consists in an improvement in the human-interaction management. Indeed, we formalized an evolution of the virtual-chain approach, developing the \textit{virtual-bike} framework, in order to emulate also a dynamical behavior on the system, in addiction to the chain equation. The proposed control law is based on the impedance control approach for robotic manipulators, see  \cite{hogan1985part1}, which aims at imposing an impedance model on the position tracking error.\\ 
For both contributions, we provide experimental results in order to validate the proposed approaches, when the rider is effectively pedaling on the bike.

The remainder of the paper is organized as follows: first of all, a vehicle overview is given in Section \ref{se_senza_model}, along with its longitudinal dynamics model. In Section \ref{se_virtual_chain}, the virtual-chain background is recalled and extended to series-parallel bikes and experimentally validated. In Section \ref{se_virtual_bike}, the development of the virtual-bike framework is described and validated through experimental tests.

\section{Vehicle overview and modeling}
\label{se_senza_model}
The experimental setup, shown in Fig. \ref{fig_senza}, is a bike equipped with a fixed-gear chain with free-wheel mechanism; a rear-drive traction motor; a mid-drive generator and a battery pack with an integrated battery management system (BMS).

\begin{figure}[b]
\centering
\includegraphics[scale=0.8]{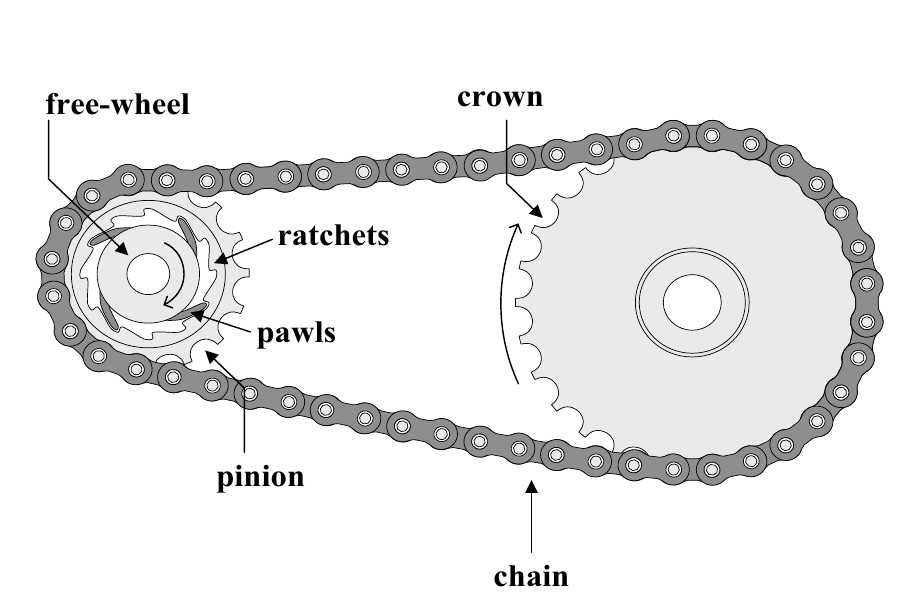}
\caption{Chain with free-wheel mechanism.}
\label{fig_chain}
\end{figure}

The key element of this vehicle is the free-wheel mechanism, shown in Fig. \ref{fig_chain}. It is placed between the rear wheel and the chain pinion and it is constituted of an external gear-wheel, composed of ratchets, and an internal gear-wheel, connected to the bike wheel. The internal one is equipped with pawls, which are pushed by springs in order to keep the internal and external gear-wheel in contact. The asymmetric shape of the ratchets makes the relative motion between the two gear-wheels possible only in one direction, i.e. when backpedaling. In this case, the chain is said to be disengaged, indeed any torque can be transmitted from the chain to the wheel. In the opposite direction, the chain is engaged transmitting the torque to the wheel only when the two gear-wheel are locked together, and this happens if and only if they are rotating at the same speed. It is remarked that the mechanical structure of the free-wheel avoids that the external gear-wheel rotates faster than the external one.\\
It follows that the free-wheel acts as a clutch and this is the reason why the vehicle can be either parallel or series. In \textit{parallel mode}, the traction power is the sum of the  motor one and the rider's one, thanks to the chain engagement, while the generator does not play any role in the traction. On the other hand, in \textit{series mode}, when chain is disengaged, the power applied at the wheel is from the motor only, while the rider's one is converted by the generator and stored in the battery, that acts as a power buffer. 

To control the vehicle, each electrical machines has a specific control unit and the generator one acts also as an overall vehicle control unit (VCU). These two boards communicate together, and with the BMS as well, through the vehicle CAN bus. Finally, the bike is equipped with mechanical brakes managed though the classical levers on the handlebar, while the regenerative braking is activated when backpedaling.

\subsection{Modeling and identification}
Given that this work is oriented to the control of the human-vehicle interaction, the model is focused on the longitudinal dynamics. Indeed, the model is derived from a force balance at the wheel:
\begin{equation}
M\frac{\mathrm{d}v}{\mathrm{d}t} = \frac{T_\mathrm{m}+T_\mathrm{br}+\xi T_\mathrm{ch}}{R_\mathrm{w}} - F_\mathrm{cd}(v),
\label{eq_senza_lon}
\end{equation}
where $M$ is the vehicle mass, that includes the bike one $M_\mathrm{b}$ and the rider's one $M_\mathrm{h}$, and $v$ is the longitudinal speed. $T_\mathrm{m}$ is the torque provided by the electric motor, $T_\mathrm{br}$ is the braking torque applied by the mechanical brakes. $T_\mathrm{ch}$ is the torque transmitted by the chain as a function of the free-wheel mechanism status $\xi$. $R_\mathrm{w}$ is the wheel radius and $F_\mathrm{cd}$ is the coasting down force. 

Concerning the free-wheel mechanism, it can modeled as a non-controllable clutch, whose status $\xi$ is a boolean value which expression is:
\begin{equation}
\left\{\begin{array}{lclccl}
\xi = 1 &\Leftrightarrow& \Omega_\mathrm{w} =\tau_\mathrm{ch}\Omega_\mathrm{p} &$ and $& T_\mathrm{ch} >0 & $chain engaged$ \\
\xi = 0 &\Leftrightarrow& \Omega_\mathrm{w} >\tau_\mathrm{ch}\Omega_\mathrm{p} &$ or $& T_\mathrm{ch} \leq 0 & $chain disengaged$ 
\end{array}\right.,
\end{equation}
where $\tau_\mathrm{ch}$ is the mechanical chain ratio.

The chain torque can be computed from a torque balance at the pedals:
\begin{equation}
J_\mathrm{g}\frac{\mathrm{d}\Omega_\mathrm{p}}{\mathrm{d}t}  = T_\mathrm{h}+T_\mathrm{g}-\tau_\mathrm{ch}T_\mathrm{ch},
\label{eq_t_chain}
\end{equation}
where $J_\mathrm{g}$ is the generator inertia. Neglecting the low inertia $J_\mathrm{g}$, \eqref{eq_t_chain} can be simplified to:
\begin{equation}
T_\mathrm{ch} = \frac{T_\mathrm{h}+T_\mathrm{g}}{\tau_\mathrm{ch}},
\label{eq_t_chain_simpl}
\end{equation}
where $T_\mathrm{h}$ is the human torque applied by the rider. 

\begin{figure}[h]
\centering
\includegraphics[scale=0.9]{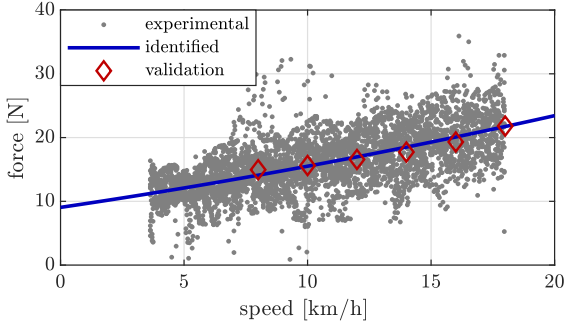}
\caption{Coasting-down force. Experimental fitting data are compared with the identified model and the validation data.}
\label{fig_senza_Fcd}
\end{figure}

Finally, the coasting-down force $F_\mathrm{cd}$ has been modeled as the sum of three different contributions, representative of the rolling resistance, the viscous and aerodynamic friction:
\begin{equation}
F_\mathrm{cd}(v) = C v^2 + B v + A,
\label{eq_senza_cd}
\end{equation}
where $A,B,C$ are model coefficients. These parameters have been identified through costing-down tests: the vehicle has been accelerated to its top speed and then left decelerate on its own. Therefore, during these tests, \eqref{eq_senza_lon} becomes:
\begin{equation}
F_\mathrm{cd}(v) = M R_\mathrm{w}^2 \frac{\mathrm{d}\Omega_\mathrm{w}}{\mathrm{d}t},
\label{eq_senza_cd_est}
\end{equation}
where $\Omega_\mathrm{w}$ is the measured angular wheel speed, while $M$ and $R_\mathrm{w}$ are known. Hence, the coasting-down coefficients have been found minimizing the error between the model, left-hand side of \eqref{eq_senza_cd_est}, and the measure, right-hand side of \eqref{eq_senza_cd_est}. Given the importance of this model in the development of the virtual-bike framework in Section \ref{se_virtual_bike}, it is validated also through tests at constant speed ($\bar{v}$), so that:
\begin{equation}
F_\mathrm{cd}(\bar{v}) = \frac{T_\mathrm{m}}{R_\mathrm{w}}.
\label{eq_senza_cd_val}
\end{equation}
In this way, the model accuracy is evaluated by comparing it with (right-hand side of \eqref{eq_senza_cd_val}). Results are shown in Fig. \ref{fig_senza_Fcd}.

In conclusion, exploiting that the motor and generator torque are proportional to the respective current, the longitudinal dynamics model can be summarized in
\begin{equation}
M\frac{\mathrm{d}v}{\mathrm{d}t} = \frac{K_\mathrm{m}I_\mathrm{m}+T_\mathrm{br}}{R_\mathrm{w}} + \xi\frac{T_\mathrm{h}+K_\mathrm{g}I_\mathrm{g}}{\tau_\mathrm{ch} R_\mathrm{w}} - F_\mathrm{cd}(v),
\label{eq_senza_model}
\end{equation}
whose the main parameters are reported in Table \ref{tab_senza_param}.\\
\begin{table}[h]
\footnotesize
\caption{Bike parameters.}
\centering
\begin{tabular}{ccc}
\hline
\textbf{parameter} & \textbf{value} & \textbf{unit} \\
\hline
\hline
$M_\mathrm{b}$ & 25 & kg \\
$R_\mathrm{w}$ & 33 & cm \\
$\tau_\mathrm{ch}$ & 1.80 & - \\
$K_\mathrm{m}$ & 1.69 & Nm/A \\
$K_\mathrm{g}$ & 10.7 & Nm/A \\
\hline
\end{tabular}
\label{tab_senza_param}
\end{table} 

\section{Virtual-chain for the human-vehicle interaction management}
\label{se_virtual_chain}
In this section, the \textit{virtual-chain} concept, proposed by \cite{corno2017senza} for series bike, is briefly recalled and then extended to series-parallel bikes.

\subsection{Virtual-chain background for series bikes}
The virtual-chain approach is the result of a master-slave bilateral control problem, see \cite{yokokohji1994bilateral}, formulated on the constitutive chain equations:
\begin{equation}
\left[\begin{array}{c}
\Omega_\mathrm{p} \\
T_\mathrm{h}\end{array}\right] = 
\left[\begin{array}{cc}
\frac{1}{\tau_\mathrm{v}} & 0 \\
0 & \tau_\mathrm{v}\end{array}\right]
\left[\begin{array}{c}
\Omega_\mathrm{l} \\
T_\mathrm{l}\end{array}\right] .
\label{chain_eq}
\end{equation}
Considering a series configuration, the load torque $T_\mathrm{l}$ and speed $\Omega_\mathrm{l}$ are the ones at the traction wheel; indeed, the wheel speed $\Omega_\mathrm{w}$ defines the vehicle speed $v$, driven by $T_\mathrm{m}$, only. Finally, $\tau_\mathrm{v}$ plays the role of a \textit{virtual-chain ratio} that can be freely designed by the user. In \cite{corno2017senza}, it is proven that in case of low generator and motor impedance, as in case of bikes, a good approximation of the bilateral control solution can be achieved with:
\begin{itemize}
\item[--] a generator cadence control with reference equal to $\Omega_\mathrm{w}/\tau_\mathrm{v}$;
\item[--] a motor torque control with reference equal to $T_\mathrm{h}/\tau_\mathrm{v}$.
\end{itemize}
To make this approach feasible, the rider's torque $T_\mathrm{h}$ knowledge is required. Therefore, it could be provided by proper pedal torque sensors or, if they are unavailable, the cyclist's torque can be estimated from the torque balance at the generator \eqref{eq_t_chain}, neglecting the generator inertia $J_\mathrm{g}$ and considering the series bike scenario $T_\mathrm{ch}=$ 0. It follows that:
\begin{equation}
T_\mathrm{h} \approx -T_\mathrm{g}.
\end{equation}
It is interesting to notice that in this situation the battery acts as a buffer, indeed all the generated power (except for the electric losses) is given to the motor:
\begin{equation}
P_\mathrm{g} = T_\mathrm{g}\Omega_\mathrm{p} = T_\mathrm{g}\frac{\Omega_\mathrm{w}}{\tau_\mathrm{v}} =
T_\mathrm{m}\Omega_\mathrm{w} = P_\mathrm{m}.
\label{eq_pg_pm}
\end{equation}

Thanks to the formulation of the bilateral control problem in \eqref{chain_eq}, this vehicle control law has the capability to enhance the experience to be on a traditional chained-bike to rider. In fact, the closed-loop system obtained by placing the virtual-chain concept in \eqref{eq_senza_model} is characterized by the following equations:
\begin{equation}
\def\arraystretch{1.2}
\left\{\begin{array}{l}
\Omega_\mathrm{p} = \frac{\Omega_\mathrm{w}}{\tau_\mathrm{v}}S_\mathrm{g}(s)\\
M\frac{\mathrm{d}v}{\mathrm{d}t} = \frac{T_\mathrm{h}}{\tau_\mathrm{v}R_\mathrm{w}}S_\mathrm{m}(s) - F_\mathrm{cd}(v)
\end{array}\right. ,
\label{eq_cl_series}
\end{equation} 
where $S_\mathrm{g}(s)$ is the transfer function of the closed-loop cadence control
and $S_\mathrm{m}(s)$ is the transfer function of the closed-loop motor torque control. It is immediately visible in \eqref{eq_cl_series}, that the vehicle actually behaves as a traditional chained-bike, when these two quantities are close to one in desired range of frequencies.

In conclusion, the advantages of this control scheme are summarized in: 1) the experience of riding a vehicle ruled by the equations of a traditional bike is enhanced; 2) the possibility of designing the virtual-chain ratio increases the versatility of this kind of vehicles, being also able to emulate more sophisticated chain strategies, up to continuously varying ratios able to maintain the pedal cadence at the cyclist's most comfortable one. However, as discussed by \cite{corno2017senza}, the solution becomes not comfortable at low bike speed; indeed, given the power motor power limits, the bilateral control problem is not achieved, i.e. $S_\mathrm{m}(s)\neq1$.

\subsection{Virtual-chain extension to series-parallel bikes}
Before introducing the extension to series-parallel bikes, the effects induced by the free-wheel mechanism, acting as an uncontrollable clutch on the virtual chain design, should be analyzed. Indeed, the presence of the mechanical chain introduces a constraint on the pedal-wheel speed plane, given that pedals can not rotate faster than the reduced wheel one. It is possible to notice that when the operating point lays on the constraint the vehicle operates in parallel configuration, otherwise in series.

\begin{figure}[h]
\centering
\includegraphics[scale=0.9]{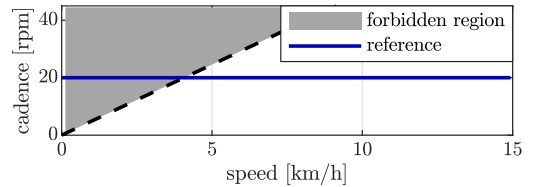}
\caption{Forbidden region in the design of the virtual chain ratio.}
\label{fig_tatio_region}
\end{figure}

To present and validate the proposed integration strategy, we used as use-case a virtual chain ratio designed to have a constant pedal cadence reference $\bar{\Omega}$. Looking at Fig. \ref{fig_tatio_region}, it is clear that at low vehicle speed the virtual reference cannot be reached because of the chain constraint. Nevertheless, at low speed the bilateral control is naturally achieved thanks to the presence of the mechanical chain itself. Therefore, to solve the bilateral control problem, in parallel mode, there is no need of a control law, so both motors do not apply any torque. While in series mode, the virtual-chain control is naturally necessary.

In this way, the longitudinal dynamics of the considered series-parallel bike, neglecting the motor torque control dynamics, can be modeled as:
\begin{equation}
M\frac{\mathrm{d}v}{\mathrm{d}t} = \xi\frac{T_\mathrm{h}}{\tau_\mathrm{v}R_\mathrm{w}} + \frac{T_\mathrm{h}}{\tau_\mathrm{ch}R_\mathrm{w}}(1-\xi) - F_\mathrm{cd}(v).
\label{eq_cl_series_parallel}
\end{equation} 
So, the bike behaves like a  traditional chained bike: 1) when  the chain is engaged ($\xi = 1$) and $\Omega_\mathrm{p} < \bar{\Omega}$, due to the mechanical structure; 2) when the chain disengages ($\xi = 0$) and $\Omega_\mathrm{p} = \bar{\Omega}$, due to the virtual-chain control. Indeed, the closed-loop systems can be written as:
\begin{equation}
\left\{\begin{array}{ll}
M\frac{\mathrm{d}v}{\mathrm{d}t} = \frac{T_\mathrm{h}}{\tau_\mathrm{ch}R_\mathrm{w}} - F_\mathrm{cd}(v), & \xi = 1\\
\\
M\frac{\mathrm{d}v}{\mathrm{d}t} = \frac{T_\mathrm{h}}{\tau_\mathrm{v}R_\mathrm{w}} - F_\mathrm{cd}(v), & \xi = 0
\end{array}\right. .
\label{eq_cl_series_parallel_2}
\end{equation} 
Looking at \eqref{eq_cl_series_parallel_2}, we can see that the closed-loop system can be considered a \textit{virtual parallel hybrid vehicle}. Indeed, when the chain is engaged the vehicle is naturally parallel, while in series mode the vehicle behaves as a chain is virtually engaged.\\
To summarize, in order to extend the virtual-chain control to a series-parallel vehicle, the control law, in principle, is a switching one. Indeed, when the chain is disengaged, the control law is the traditional virtual-chain; while, when the chain is engaged, the two electrical machines need to be tuned-off, i.e. apply zero current. However, if the generator is properly controlled to have zero current, the virtual-chain motor control law $T_\mathrm{m} = T_\mathrm{g}/\tau_\mathrm{v}$ has already an outcome equal to zero; therefore, the only switching control law present on the vehicle is at the generator side. In conclusion, the generator voltage control law is defined so to apply: 1) zero current when the cadence is lower than the virtual-chain cadence reference; 2) control the cadence  to keep desired reference once it is reached. It is possible to notice that, according to this principle, also if the series mode is active but the cadence is lower than the reference the generator applies zero current. Hence, the pedals cannot push the rider to the reference, who instead feel a zero load torque, when is pedaling slower than the reference cadence. This results to be an emulation of the free-wheel mechanism, in series mode.

The switching between the two controllers, in practice, could become uncomfortable, e.g. because of to measurement noise and the oscillatory behavior of the rider's pedaling itself. Therefore, to merge the controllers in a smoothed way, we followed the methodology proposed by \cite{radrizzani2021concurrent}, applying the minimum between the two control actions. Indeed, when the cadence is lower than the desired one, the cadence controller requires a positive current to accelerate the pedal, while the current controller asks for lower voltage aiming at zero current. On the other hand, when the desired cadence is reached, the cadence controller must ask for a negative torque to avoid the overcome of the reference; opposite, the current controller desires higher voltage to maintain zero current. Finally, an anti-windup scheme for such an integration strategy is also discussed in \cite{radrizzani2021concurrent}.

\subsection{Experimental validation}
\begin{figure}[h]
	\centering
	\includegraphics[scale=0.86]{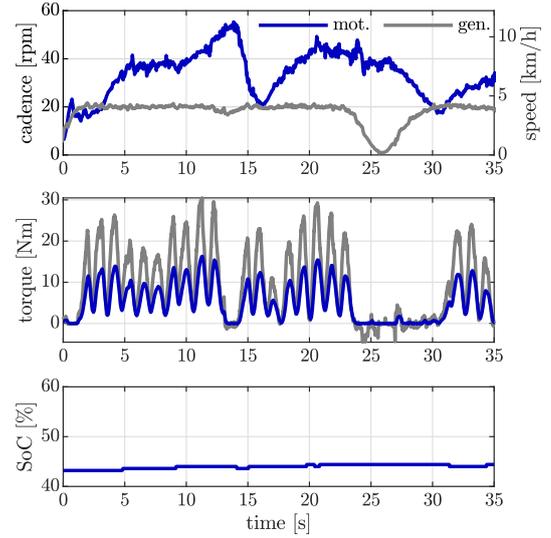}
	\caption{Validation of the virtual-chain extension to series-parallel bikes. The virtual-chain ratio $\tau_\mathrm{v}$ is chosen so to have a pedal cadence of 20 rpm. In the first plot the generator speed is compared to the motor one. Motor speed is scaled to the pedal through $\tau_\mathrm{ch}$ on the left axis and converted to vehicle speed on the right axis.}
	\label{fig_virtual_chain_val}
\end{figure}

In this section, an experimental validation of the virtual-chain extended to series-parallel bikes is given. To this aim, the virtual-chain ratio is designed so to have a fixed pedal cadence $\bar{\Omega}$ equal to 20 rpm. The effectiveness of the proposed solution is proven in Fig. \ref{fig_virtual_chain_val}: 
\begin{itemize}
\item[1)] in the first phase of the test (0-3.5 s), the bike starts moving from a stopped position, therefore the vehicle is in parallel mode and no torque is applied by the motor and the generator; 
\item[2)] at 13-14 s and 23-39 s, it is possible to see that the free-wheel mechanism in properly emulated in series configuration, since motor and generator torques are both equal to zero while the rider slows down the pedals;
\item[3)] in the other cases, the vehicle is in series mode, operating according to the virtual-chain control law. Indeed, the pedal cadence is kept at the reference while the vehicle continues to accelerate thanks to the electric motor torque applied.
\end{itemize}
It is finally shown that the battery state of charge remains almost constant during the maneuver; indeed, as discussed in \eqref{eq_pg_pm}, all the power generated by the cyclist is transmitted to the wheel by the motor. 

\section{Virtual-bike for the human-vehicle interaction management}
\label{se_virtual_bike}
In this section, we show how we extended the virtual-chain approach to a \textit{virtual-bike} framework, in order to impose a dynamical behavior in addition to the static chain model. Toward this goal, the background of impedance control defined by \cite{hogan1985part1} for robotic manipulator is recalled first to be applied as control strategy to impose the virtual-bike model. 

\subsection{Impedance control background}
The definition of impedance refers to the ratio of the 
output effort and the input flow [\cite{song2017impedancereview}], i.e. in case of manipulators, the ratio between the 
contact force resulting from the interaction motion between manipulator and environment and its velocity. Therefore, to assign an impedance model means to impose the dynamical model that links the manipulator velocity and the external acting force. 

Considering the case of manipulators, the impedance control law is derived as follows. First of all, a mathematical model of the manipulator needs to be defined:
\begin{equation}
M(x) \ddot{x} + C(x,\dot{x}) \dot{x} + g(x) = J^{-\mathrm{T}}T - F_\mathrm{ext},
\label{eq_manip_model}
\end{equation}
where $x$ is the vector representing the position and 
orientation of the manipulator; $M$, $C$, $g$ are the mass matrix, the Coriolis matrix and the 
gravity force, respectively; $J$ is the Jacobian that relates the joint with the end-effector 
velocity; and $F_\mathrm{ext}$ is the external force. Finally, $T$ is the controllable vector of forces imposed by actuators. \\
The second step is the imposition of an impedance model, usually defined in this context as a second order linear model between the external force and the position error:
\begin{equation}
M_\mathrm{v} (\ddot{x}-\ddot{x}_\mathrm{ref}) + B_\mathrm{v} (\dot{x}-\dot{x}_\mathrm{ref}) + K_\mathrm{v} ({x}-{x}_\mathrm{ref}) = -F_\mathrm{ext},
\label{eq_manip_virtual}
\end{equation}
where ${x}_\mathrm{ref}$ is the vector of the desired position and $M_\mathrm{v}$, $B_\mathrm{v}$, $K_\mathrm{v}$ are the virtual model parameters to impose.\\
The solution of the problem is given by the computation of $T$ in \eqref{eq_manip_model} such that the closed-loop systems behaves as in \eqref{eq_manip_virtual}. The result of this computation is
\begin{equation}
\def\arraystretch{1.2}
\begin{array}{ll}
T = & J^\mathrm{T} \Big[ F_\mathrm{ext} + M(x) \ddot{x} + C(x,\dot{x}) \dot{x} + g(x) +\\
& + M(x)\left(\ddot{x}_\mathrm{ref} - M^{-1}_\mathrm{v} \left( B_\mathrm{v} (\dot{x}-\dot{x}_\mathrm{ref}) + K_\mathrm{v} ({x}-{x}_\mathrm{ref}) + F_\mathrm{ext} \right) \right) \Big].
\end{array}
\label{eq_manip_res}
\end{equation}
It can be easily seen that substituting this expression of $T$ in \eqref{eq_manip_model} returns exactly \eqref{eq_manip_virtual}. It is clear that this result is also achieved on the real system when the manipulator model is accurate and the external force is measurable. 

\subsection{Virtual-bike formulation}
To develop the \textit{virtual-bike} control strategy, the same steps of the impedance control are followed. First of all, we defined the vehicle model by linearizing the longitudinal dynamics in \eqref{eq_senza_lon}, considering the bike in series configuration:
\begin{equation}
M\frac{\mathrm{d}v}{\mathrm{d}t} = \frac{T_\mathrm{m}}{R_\mathrm{w}} - \beta v, 
\end{equation}
where $\beta$ is a function of the coasting-down model introduced in \eqref{eq_senza_cd}, i.e.:
\begin{equation}
\beta = \beta(\bar{v}) = \frac{\partial F_\mathrm{cd}}{\partial v} \Big|_{\bar{v}}  = 2C\bar{v} + B.
\end{equation}
The virtual-bike is constructed over the virtual-chain, therefore, introducing the scaling factor $\kappa$ used as control variable:
\begin{equation}
M\frac{\mathrm{d}v}{\mathrm{d}t} = \kappa\frac{T_\mathrm{h}}{\tau_\mathrm{v}R_\mathrm{w}} - \beta v, 
\label{eq_senza_lin}
\end{equation}
where $T_\mathrm{h}$ plays the same role of $F_\mathrm{ext}$ in manipulators.\\
The second step is the definition of an impedance model, representative of the virtual-bike to be imposed. For example, we can define a first-order linear system:
\begin{equation}
M_\mathrm{v}\frac{\mathrm{d}v}{\mathrm{d}t} = \frac{T_\mathrm{h}}{\tau_\mathrm{v}R_\mathrm{w}} - \beta_\mathrm{v} v, 
\label{eq_senza_virtual}
\end{equation}
where $M_\mathrm{v}$ and $\beta_\mathrm{v}$ represents the parameters of the desired bike.

Given the linearity of the vehicle model \eqref{eq_senza_lin} and the virtual one \eqref{eq_senza_virtual}, the solution of the impedance control can be easily derived in Laplace domain. Indeed, introducing the Laplace operator $s $, we can compute $\kappa$ such that the following equations hold:
\begin{equation}
\begin{array}{lcl}
(Ms+\beta)v = \kappa \frac{T_\mathrm{h}}{\tau_\mathrm{v} R_\mathrm{w}} & $and$ &
(M_\mathrm{v}s+\beta_\mathrm{v})v = \frac{T_\mathrm{h}}{\tau_\mathrm{v} R_\mathrm{w}}
\end{array}.
\end{equation} 
The solution is obtained by dividing the previous two equations:
\begin{equation} 
\kappa = \frac{Ms+\beta}{M_\mathrm{v}s+\beta_\mathrm{v}}.
\end{equation} 

Finally, the complete electric motor control law for $T_\mathrm{h}\neq0$, in the virtual-bike framework becomes:
\begin{equation} 
T_\mathrm{m} = \frac{Ms+\beta}{M_\mathrm{v}s+\beta_\mathrm{v}}\frac{T_\mathrm{h}}{\tau_\mathrm{v}}.
\end{equation}
This result can be interpreted as a dynamical scaling factor $\kappa$ with the following features:
\begin{equation}
\begin{array}{lcl}
$at steady-state $ (s \rightarrow 0) &\Rightarrow & \kappa \rightarrow \frac{\beta}{\beta_\mathrm{v}}\\
$at high-frequency $ (s \rightarrow \infty) &\Rightarrow & \kappa \rightarrow \frac{M}{M_\mathrm{v}}\\
\label{eq_vb_extremes}
\end{array}.
\end{equation}
It follows that, the scaling factor can be designed so to have at steady-state different properties with respect to high-frequency. For example, to assist the rider during accelerations, we can choose $M_\mathrm{v} < M$, while recharging the battery at constant speed imposing $\beta_\mathrm{v} > \beta$. It is trivial to see that the solution and the virtual-chain coincide when $M_\mathrm{v} = M$ and $\beta_\mathrm{v} = \beta$, at the same time. \\
As in case of manipulators, the system model must be accurate and the external force measurable. In our case, the cyclist's torque is a measured quantity, as previously discussed, and to have a model of the vehicle as accurate as possible the cyclist's weight is considered as an input provided by the cyclist as well. The vehicle parameters could be estimated in real-time, to provide more robust results, e.g. as shown in \cite{lu1991adaptive} for standard impedance control.  

In the following section, the experimental validation of the proposed virtual-bike framework is provided. 

\subsection{Experimental results}
To validate the virtual-bike approach, different experimental tests have been carried out, for different values of the virtual bike parameters, summarized in Table \ref{tab_vb_test}. 

\begin{table}[h]
\footnotesize
\centering
\caption{Vehicle and virtual parameters ratio \\ when $\bar{v} = $ 6.5 km/h.}
\begin{tabular}{cccc}
\hline
\textbf{parameter} & \textbf{test 1} & \textbf{test 2} & \textbf{test 3}\\
 & \textbf{light} & \textbf{medium} & \textbf{heavy}\\
\hline
\hline
${M}/{M_\mathrm{v}}$ & 1.47 & 1.02 & 0.68 \\
${\beta(\bar{v})}/{\beta_\mathrm{v}} $ & 2.10 & 1.26 & 0.70 \\
\hline
\end{tabular}
\label{tab_vb_test}
\end{table}

\begin{figure}[b]
	\centering
	\includegraphics[scale=0.88]{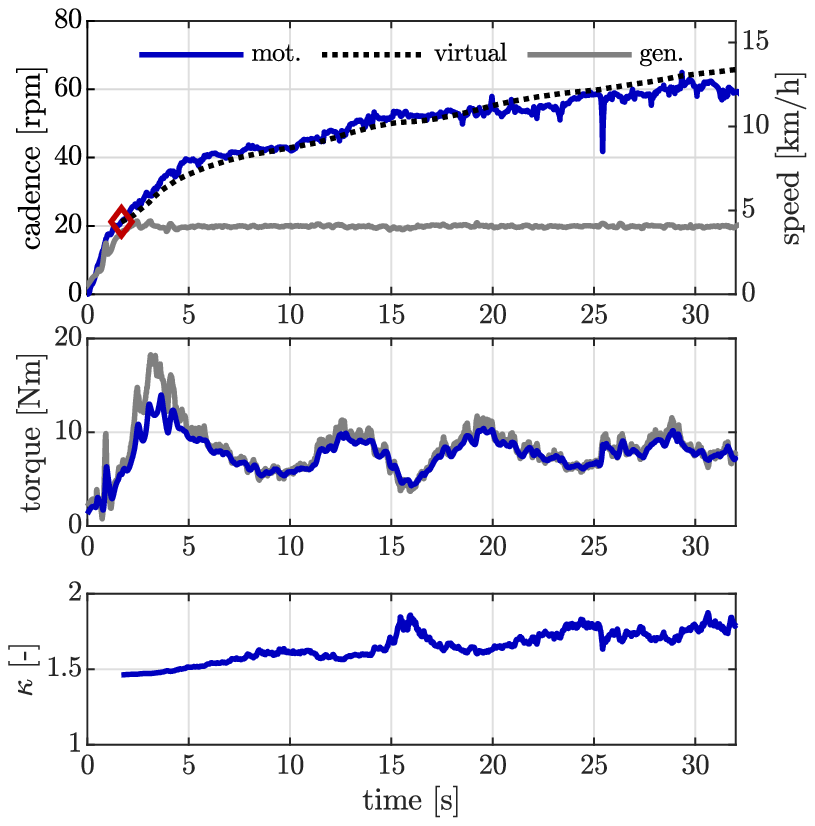}
	\caption{Test 1. Vehicle speed tracks the desired virtual-bike model, while the rider's cadence is kept at 20 rpm, by the virtual-chain ratio $\tau_\mathrm{v}$. Virtual-bike is designed as a heavy bike, so the assistance factor $\kappa$ is always greater that 1.}
	\label{fig_vb_test1}
\end{figure}

In the first test, the virtual-bike parameters are designed to have a lighter bike with less resistance. The experimental data of this test are shown in Fig. \ref{fig_vb_test1}, where torques and speeds are notch-filtered to clean signals from the cyclist disturbance, as for other tests. First of all, the good matching between the measured vehicle speed and the virtual speed provides a general validation of the proposed approach. The virtual speed is computed by feeding the virtual desired model with the measured cyclist's torque, initialized when the chain disengages, corresponding to the red diamond marker in the plots. Moreover, the time evolution of the scaling factor $\kappa$ provides a second validation. Indeed, it is always greater than 1 in accordance to the necessity of giving an assistance to the rider both in acceleration and steady-state to experience a lighter and less resistant bicycle.

\begin{figure}[b]
	\centering
	\includegraphics[scale=0.88]{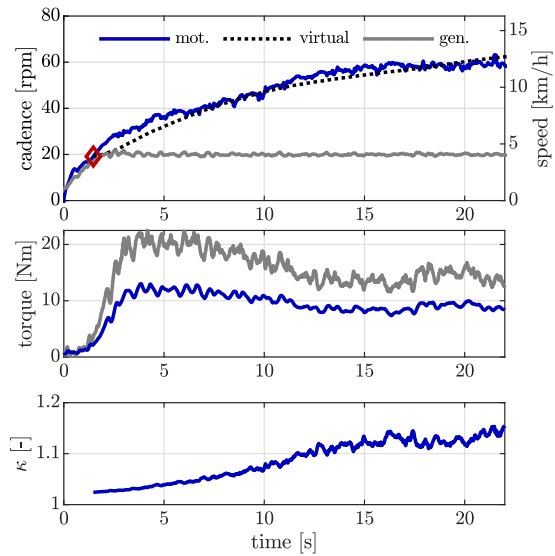}
	\caption{Test 2. Vehicle speed tracks the desired virtual-bike model, while the rider's cadence is kept at 20 rpm, by the virtual-chain ratio $\tau_\mathrm{v}$. Virtual-bike is designed as a heavy bike, so the assistance factor $\kappa$ is always close to 1.}
	\label{fig_vb_test2}
\end{figure}

In the second test in Fig. \ref{fig_vb_test2}, again the good matching between the measured speed and the ideal one is highlighted. Then, it is visible how in the acceleration phase $\kappa$ is closer to 1 and then increases at steady-state in accordance with the virtual parameter listed in Table \ref{tab_vb_test}. 

\begin{figure}[h]
	\centering
	\includegraphics[scale=0.88]{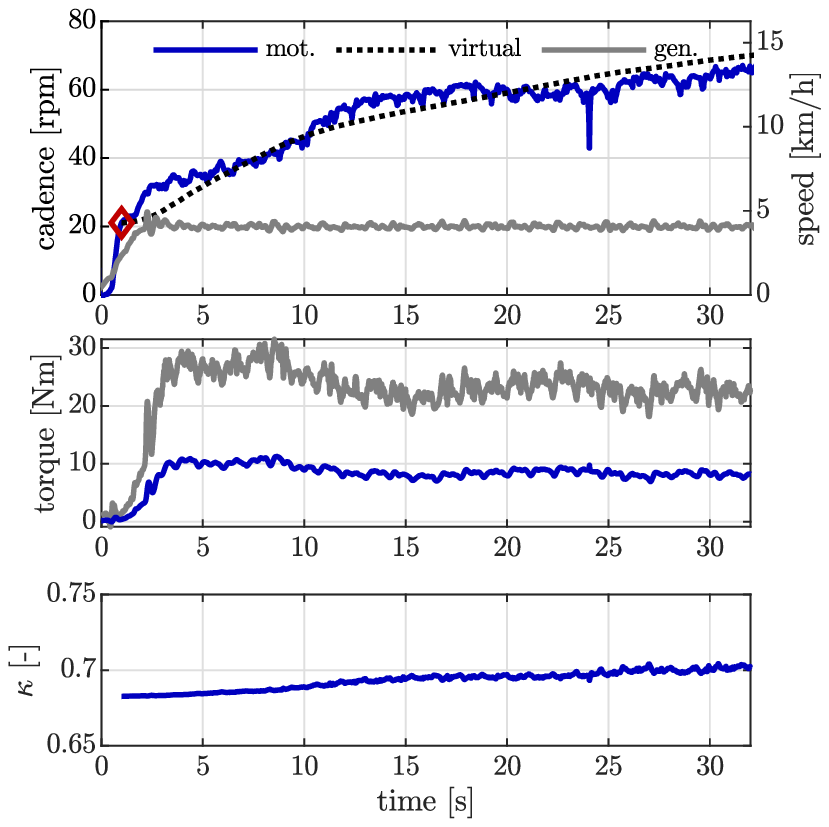}
	\caption{Test 3. Vehicle speed tracks the desired virtual-bike model, while the rider's cadence is kept at 20 rpm, by the virtual-chain ratio $\tau_\mathrm{v}$. Virtual-bike is designed as a heavy bike, so the assistance factor $\kappa$ is always lower that 1.}
	\label{fig_vb_test3}
\end{figure}

Similar results are obtained also in the third test in Fig. \ref{fig_vb_test3}. The measured speed matches the desired one thanks to value of $\kappa$ always lower than one. This implies that the rider should provide a higher torque to ride the vehicle at similar speeds of other tests, both in traction and steady-state phases.

These experimental results showed the effectiveness of the virtual-bike in emulating the dynamics of different bikes, thanks to the series mode. While the parallel mode active at low speed makes the rider experience a good feeling in starting.

\section{Conclusions}
In this work, we presented how to use the virtual-chain framework, native for series bikes, in series-parallel ones. After having experimentally validated the approach, it becomes the base to develop a virtual-bike framework: a control law able to make the rider experience a virtual bike, whose parameters are user-chosen. The proposed approach has been validated through an experimental campaign. Possible future steps are the improvement of the control law, by self-adapting the rider-depending parameters, and the virtual-bike parameters selection in an energy management perspective.

{\fontsize{9pt}{9pt}\selectfont \bibliography{library}}             

\end{document}